# HIGH PERFORMANCE TEMPERATURE CONTROLLER: APPLICATION TO THE EXCESS NOISE MEASUREMENTS OF YBCO THERMOMETERS IN THE TRANSITION REGION

B. GUILLET, L. MÉCHIN AND D. ROBBES

*GREYC -CNRS UMR 6072, ENSICAEN et Université de Caen,*
*6 Bd Maréchal Juin, 14050 Caen cedex, France*
*E-mail: bguillet@greyc.ismra.fr*

Dedicated read-out electronics was developed for low impedance resistive thermometers. Using this high performance temperature controller, the temperature dependence of the excess noise of a $YBa_2Cu_3O_{7-\delta}$ (YBCO) sample in the superconducting transition was monitored as a function of the current bias. The noise could reach $3 \times 10^{-8}$ K $Hz^{-1/2}$ at 1 Hz, 5 mA bias and 90 K.

## 1. Introduction

Temperature fluctuations are a limiting factor for absolute thermometry, bolometry and radiometry [1] or more generally for sensitive instruments operating at low temperature. There is a need for sensors and systems that are able to precisely measure and control the temperature, respectively. The low-frequency 1/f noise in high-Tc superconducting (HTS) thin film sensors have been studied at 300 K as function of the deposition conditions, the microstructure [2]. However such physical correlations are more difficult to address in the normal-superconducting transition temperature range, which is the high sensitivity domain for such sensors. At the transition the temperature coefficient dR/dT can reach few 100 $\Omega$ $K^{-1}$ and temperature fluctuations caused by heating power instabilities can significantly contribute to the measured noise. The dedicated system presented here uses two sensitive HTS film in a closed loop and performs a large rejection of these parasitic heat perturbations. It therefore enables to properly study the intrinsic excess noise of HTS films at the superconducting transition for different current bias conditions.

## 2. Characteristics of the thin film HTS thermometers

The studied thin film HTS thermometers consist in a high quality 200 nm thick $YBa_2Cu_3O_{7-\delta}$ (YBCO) film deposited on a $SrTiO_3$ substrate by pulsed laser deposition. Samples are patterned in 40 μm×600 μm strips with four gold contacts. The homogeneity of the deposited film is assessed by the small dispersion in the critical temperature $T_c$ extracted from the resistance versus temperature plots and their derivatives for a constant current bias of 1 mA. A dimensionless characteristic slope of these thermometers, defined as A=dlnR/dlnT, is around 150 comparable to values found in literature [3].

## 3. Read-out electronics and experimental setup

A dedicated read-out electronics was developed for these low impedance resistive thermometers (<10 k$\Omega$) [4]. The system consists in a highly stable square modulated current source, a low-noise preamplifier and an integrated lock-in amplifier coupled to an optimized analog PID controller (see Fig. 1). This system was originally designed to monitor an active cavity radiometer operating at 90 K in a liquid nitrogen bath.

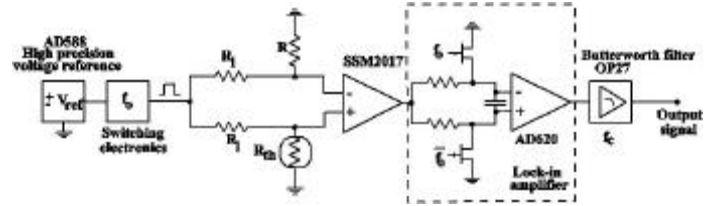

Fig 1. Home made read-out electronics developed for any low impedance resistive thermometer [4].

The bias voltage is applied to two high precision dividers : one includes the resistive thermometer $R_{th}$ and the other includes a low Thermal Coefficient Resistance (TCR) R. In order to limit self-heating effects, the electrical power dissipated in the thermometer as well as in the low TCR resistor R is below 1 mW. The preamplifier stage is based on a SSM2017 integrated instrumentation amplifier followed by a lock-in amplifier. A second-order Butterworth filter shapes the heating output signal of the battery-operated power device. The white noise at the switching frequency $f_o$ is mainly due to the Johnson noise of the resistances R and $R_{th}$. At the operating temperature, the value of R is chosen close to $R_{th}$, we can therefore estimate the voltage equivalent noise $e_n=(8k_BTR)^{1/2}$. Including the preamplifier input noise, we obtain the equivalent noise source $e_n \sim 2$ nV Hz$^{-1/2}$ at room temperature and R=100 $\Omega$. To evaluate the total noise of the read-out electronic system, the thermometer was replaced by a second low TCR resistor (Fig. 2).

The system exhibits a close "1/f" low frequency noise type below 1Hz [4]. Results assessing the temperature regulation performances on copper plates, 3-cm diameter and 1 mm thick, in vacuum and in electrically and thermally shielded environment, using (YBCO) transition edge sensor (TES) were presented earlier [5]. The good low frequency performances of the setup (see Fig. 2) and the thermal stability of the sample holder (thermal drift <0.2 µK s$^{-1}$ at 90K) make sensitive temperature measurements possible. From such sets of measurements in closed loop, the efficiency of the heat perturbation rejection was estimated to be 400 at dc, in a bandwidth of about 50 mHz. It should be noted that this system can be extended to larger temperature ranges if used with selected thermometers at each temperature range. The read-out electronics is flexible enough to enable change in the bias current amplitude (with several order of magnitude). The spectral range covered by the system may be as large as 5 orders of magnitude up to 100 Hz.

In our previous study [5], the YBCO thermometer was biased using a square modulated current bias at a frequency of 1 kHz and an amplitude of 0.83 mA. In these conditions, the Noise Equivalent Temperature (NET) of the read-out electronics with an ideal YBCO thermometer (*i.e.* without excess noise), defined as the ratio between the noise $S_v$ and the responsivity of the thermometer at 1 Hz, would be ranged from 5 to 20 nK at frequencies above a few Hz. However NETs obtained from experimental results were two orders of magnitude higher indicating that the contribution of excess noise sources at the superconducting transition is important (see Fig. 2).

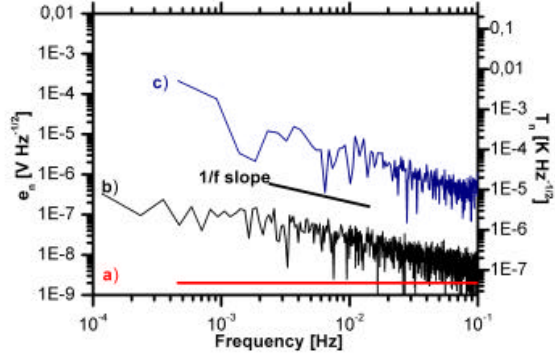

Fig. 2. Input noise spectral density $e_n$ [V Hz$^{-1/2}$] and Noise Equivalent Temperature (NET) $T_n$ [K Hz$^{-1/2}$] for the read-out electronic system without thermometer (curve b) and with the YBCO thermometer (0.83 mA bias) (curve c) as well as the theoretical white noise value (curve a) are plotted [4].

To further reduce the NET, we found necessary to study the mechanisms of the low frequency excess noise of YBCO sample operated in the superconducting transition. In that particular case, the semi empirical model of Hooge is not applicable as the normal conducting charge carriers number as well as the density of Cooper pairs vary in this temperature range. In the experiments presented here, a first YBCO sample was used as the in-loop thermometer, as in the previous configuration, in order to provide the highly stable temperature while the voltage fluctuations $S_v$ of the other YBCO sample were monitored as a function of the bias current and temperature.

## 4. Results and discussions

At a low current bias of 0.75 mA, as indicated previously for 0.83 mA, a large excess noise was measured in the superconducting transition region. A voltage noise power spectral density peak appeared at the maximum of dR/dT (see Fig. 3). Similar voltage fluctuation peaks in the transition region have been already observed. Several physical mechanisms have been suggested to interpret these noises sources:
- variations of the superconducting phase fraction when the current is passing through the film [6],
- temperature fluctuations [7],
- phonon noise of the film connected to the heat sink via the thermal conductance.

However at low current bias, the normalized noise $S_v/V^2$ increased with decreasing temperature probably pointing out a percolation regime [6]. Whereas at higher current bias (5 mA) the peak associated to the superconducting transition was not observed. As $S_v/V^2$ decreased with decreasing temperature, the film is probably in the "bulk" regime [6].

In order to optimize the temperature measurements in our control system, the NET has been calculated for each bias current at the temperature, where the responsivity is maximum. It decreases with increasing bias current (noise decreases and responsivity increases) (see Fig. 4). At large current bias, the voltage noise spectral density should reach the Johnson noise.

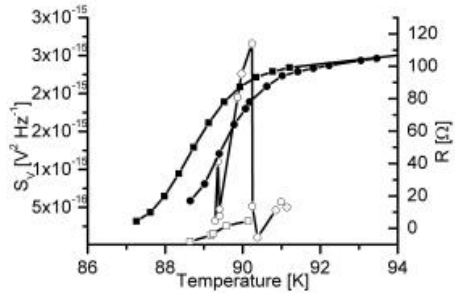 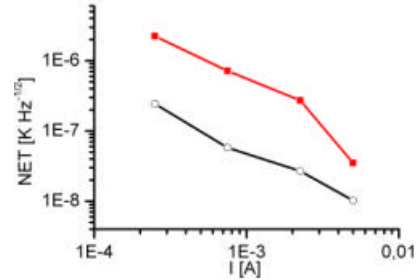

Fig. 3: Resistance R (closed symbols and right axis) and the excess noise $S_v$ (1 Hz) (opened symbols and left axis) for 0.75 mA (circles) and 5 mA (squares) in the transition region.

Fig. 4: Noise Equivalent Temperature (NET) at 1 Hz of the YBCO sample versus the current bias: the measured value (closed symbols) and the theoretical Johnson value (opened symbols)

However, there is a practical limiting value of the bias current in order to avoid thermal runaway that would deteriorate the thermometer. Measurements at 5 mA bias current amplitude improved our previously reported configuration (0.83 mA) by more than an order of magnitude. The achieved NET was $3\times10^{-8}$ K Hz$^{-1/2}$ at 1 Hz and 90 K similar to best reported values [3].

**Acknowledgements**

The authors would to thank K. Mercha for fruitful discussions.